\documentclass[10pt,letterpaper,aps,prl,reprint,superscriptaddress,showpacs,
twocolumn,nofootinbib,floatfix,amsmath,amssymb]{revtex4-1}

\usepackage{amsmath}
\usepackage{amsfonts}
\usepackage{amssymb}
\usepackage{mathrsfs}
\usepackage{braket}

\usepackage{color}
\usepackage{hyperref}

\DeclareMathOperator{\tr}{tr}

\begin{document}

\title{Quantum gravity tomography}

\author{William Donnelly}
\affiliation{
Perimeter Institute for Theoretical Physics, 31 Caroline St. N. Waterloo, ON, N2L 2Y5, Canada}
\email{wdonnelly@perimeterinstitute.ca}

\begin{abstract}
The holographic principle posits that all quantum information in a region of spacetime is encoded on its boundary. While there is strong evidence for this principle in certain models of quantum gravity in asymptotically anti-de Sitter spacetime, it is yet to be established whether holography is a generic feature of quantum gravity, or a peculiar property of these models. The goal of the present work is to present a model of holographic reconstruction in the framework of perturbative quantum gravity in flat spacetime. Specifically, we consider a state in the single-particle sector of a quantum field theory and give a method to completely reconstruct the quantum state from measurement of the metric at spatial infinity. Our argument uses a relativistic generalization of the quantum-mechanical Wigner function, and gives an explicit mechanism by which the gravitational constraints encode quantum information holographically on the boundary. Moreover, it suggests how information about more general states might be recovered from soft charges at null infinity, with applications to the black hole information loss paradox.
\end{abstract}

\maketitle

\section{Introduction}

The holographic principle is the conjecture that all information in a quantum gravitational system is encoded on its boundary \cite{tHooft:1993dmi,Susskind:1994vu}.
This principle is most concretely realized in gauge/gravity duality, where information about quantum gravity in asymptotically anti de Sitter spacetime is encoded in a conformal field theory on its asymptotic boundary.
An essential feature of gravity underlying the holographic principle is that the Hamiltonian of general relativity is a surface integral \cite{Arnowitt:1962hi}.
This implies boundary unitarity \cite{Marolf:2008mf}: observables localized on the boundary at one time remain accessible on the boundary at all times.
Boundary unitarity underlies the resolution of the black hole information paradox in holographic theories \cite{JACOBSON:2013ewa}.

The essential question then becomes not how unitarity is preserved, but rather how local bulk physics is encoded on the boundary.  
In gauge/gravity duality, much progress has been made on the question of bulk reconstruction \cite{DeJonckheere:2017qkk,Harlow:2018fse}.
Existing methods for bulk reconstruction \cite{Hamilton:2006az} are based on expansion of the free field equations, and are therefore tied essentially to the asymptotically anti-de Sitter boundary conditions.
Paradoxically, these methods do not make any explicit use of gravitational physics in the bulk.
This raises the question of how well the holographic principle may generalize to other settings, such as quantum gravity in asymptotically flat spacetime.

In the language of quantum information theory, the question is whether the set of observables on the boundary of Minkowski space is \emph{tomographically complete}, that is whether they are sufficient to completely determine the state in the bulk.

An alternative approach to the problem of bulk reconstruction makes direct use of global charges.
The fact that the total energy is a surface integral allows one to distinguish states of different total energy \cite{Balasubramanian:2006iw}.
This is not sufficient to recover a quantum state that is a superposition of energy eigenstates; for such a state one requires a complementary observable that does not commute with the energy.
In asymptotically flat spacetime, not only the energy but all Poincar\'e charges are surface integrals \cite{Regge:1974zd}.
Ref.~\cite{Chatwin-Davies:2015hna} used this fact to give a protocol for extracting quantum information encoded in the spin of a particle behind a black hole horizon.
The recovery of quantum information hinges on the fact that one can measure all three components of the angular momentum via the asymptotic charges; these are known to be tomographically complete for a spin-$\frac{1}{2}$ system.

In this letter we show that the algebra generated by the Poincar\'e charges, all of which are measurable at infinity, are tomographically complete for single particle states.
The key observation is that in quantum gravity the momentum and center of energy are a complementary pair of global charges that form a natural basis for tomography.
We will show that the observables generated by these two operators are tomographically complete, and give a procedure for reconstructing the quantum state of a single particle from the asymptotic gravitational field.
This gives an explicit mechanism for how the gravitational constraints ``imprint'' quantum information about a bulk state of matter onto the asymptotic gravitational field \cite{JACOBSON:2013ewa}.
This mechanism assumes nothing beyond properties of perturbative quantum gravity in flat spacetime.

\section{Perturbative gravity}

Consider a free scalar field theory in $(d+1)$-dimensional Minkowski spacetime.
The field admits a mode expansion (following \cite{Srednicki:2007qs})
\begin{equation}
\hat \phi(x) = \int \widetilde{dk} \, (\hat a_k e^{i k \cdot x} + \hat a_k^\dagger e^{- i k \cdot x}), \qquad \widetilde{dk} := \frac{d^d k}{(2 \pi)^d} \frac{1}{2 \omega_k}.
\end{equation}
We will consider a general mixed state of a single particle
\begin{equation} \label{rho}
    \hat \rho = \int \widetilde{dk} \, \widetilde dk' \; \rho(k,k') \ket{k} \! \! \bra{k'}
\end{equation}
where $\ket{k} = \hat a_k^\dag \ket{0}$.

We now perturbatively couple this theory to gravity by considering the perturbed Minkowski metric
$g_{\mu \nu} = \eta_{\mu \nu} + h_{\mu \nu}$.
We will work in the linearized approximation where we treat $h_{\mu \nu}$ as a small perturbation and neglect terms of order $h^2$ and higher.
While the difficulties of quantizing gravity nonperturbatively are well known, the theory can be quantized perturbatively about flat spacetime \cite{Feynman:1963ax}.

The essential feature of the quantum theory for our purposes is that the constraint equations hold in matrix elements between physical states:
\begin{equation} \label{constraint}
\bra{\phi} \left( \hat G_{0 \mu} - 8 \pi G \hat T_{0 \mu} \right) \ket{\psi} = 0,
\end{equation}
where $\hat G_{\mu \nu}$ is the linearized Einstein tensor, a linear function of the second derivative of the quantized metric perturbation $\hat h_{\mu \nu}$, and 
$\hat T_{0 \mu}$ is a component of the (suitably regularized) stress-energy tensor operator.

The condition \eqref{constraint} is a restriction on allowed physical states of quantum gravity.
Starting from a quantum state in a fixed background such as $\hat \rho$ \eqref{rho}, it can be ``gravitationally dressed'' to obtain a state that satisfies the constraint and reduces to the original state to leading order in a perturbative expansion. Perturbative gravitational dressing of operators was considered in Refs.~\cite{Donnelly:2015hta,Donnelly:2016rvo,Donnelly:2017jcd} and generalized to states in Ref.~\cite{Donnelly:2018nbv}.
Since the constraint acts on both the gravity and matter sector, dressed states are necessarily entangled states of matter and gravitational degrees of freedom.
There are many possible ways to dress a given state, corresponding to the freedom to add gravitons to the state, including ``soft gravitons''.
Our argument will be independent of the choice of dressing, we will assume that an appropriate gravitational dressing of $\hat \rho$ has been chosen.

When $\xi$ is a Killing vector of Minkowski spacetime we can integrate the constraint \eqref{constraint} to obtain a global charge.
The generator of an infinitesimal transformation $\xi$ is
\begin{equation}
\hat H_\xi := \int d^d x \, \hat T_{0\mu}(x) \, \xi^\mu(x) = \oint dS \; \hat h_\xi(x),
\end{equation}
where the last integral is over the $(d-1)$-sphere at spatial infinity.
The density $\hat h_\xi(x)$ is linear in $\hat h_{\mu \nu}$ and its first derivatives; explicit expressions are given in Refs.~\cite{Regge:1974zd,Donnelly:2016rvo}.

These global charges can be used to extract information about the matter state using the asymptotic gravitational field.
Using the gravitational constraints, an expectation value of any product of global charges is given by a product of surface integrals:
\begin{equation}
\langle \prod_{i} \hat H_{\xi_i} \rangle_\rho = \langle \prod_i \oint dS \, \hat h_{\xi_i}(x) \rangle_\rho.
\end{equation}
Thus correlation functions of global charges can be expressed in terms of the $n$-point function of the metric, evaluated at spatial infinity.

We will be interested in the case when $\xi$ generates a translation or Lorentz boost.
The generator of translations in the $i^\text{th}$ direction is the momentum
\begin{equation}
\hat P^i = \int d^d x \; \hat T^{0i}(x),
\end{equation}
and the generator of a Lorentz boost about the origin along direction $i$ is the first moment of the energy density:
\begin{equation}
    \hat K^i = \int d^d x \; x^i \; \hat T^{00}(x).
\end{equation}
Their commutator is
\begin{equation} \label{KPcommutator}
[\hat K^i, \hat P^j] = i \, \hat P^0 \, \delta^{ij}.
\end{equation}
Thus $\hat K$ provides a natural generalization of position in the relativistic theory: in the nonrelativistic limit $\hat K^i \to m \hat x^i$ and $\hat P^0 \to m$, \eqref{KPcommutator} becomes the canonical commutation relation $[\hat x^i, \hat p^j] = i \delta^{ij}$.
This reflects the fact that in this limit, a Lorentz boost with generator $\hat K^i$ becomes a Galilean boost with generator $m \hat x^i$.

We are now interested in the question of how to reconstruct the density matrix from measurements of asymptotic charges $\hat K$ and $\hat P$.
Before addressing this question it is useful to first consider the nonrelativistic limit, in which the problem becomes that of recovering the density matrix of a nonrelativistic particle from measurements of $\hat x$ and $\hat p$.
This is a well-known question in quantum state tomography, which we will now review as a prelude to generalization to the relativistic case.

\section{Wigner function}

We give a brief summary of the Wigner function \cite{Wigner:1932eb} in quantum mechanics and its application to quantum state tomography. 
We work in one dimension for simplicity, but the generalization to the multidimensional case is straightforward.

The Wigner function $w(x,p)$ is a real-valued function on phase space which encodes all the information present in the density operator.
One defining property of the Wigner function is that its moments encode expectation values of symmetrically-ordered monomials in $\hat x$ and $\hat p$.
Specifically, let $\sigma(\hat x^m \hat p^n)$ denote the symmetric ordering of the monomial $\hat x^m \hat p^n$.
The expectation value of this operator can be obtained from the Wigner function as if the latter were a classical distribution:
\begin{equation} \label{moments}
\langle \sigma( \hat x^m \hat p^n) \rangle = \int dx \, dp \; w(x,p) \; x^m p^n.
\end{equation}
Thus $w$ acts in this respect like a probability distribution on phase space, with the notable distinction that $w(x,p)$ can be negative.

The property \eqref{moments} defines $w$ uniquely.
To see this, it is useful to first define the \emph{characteristic function} \cite{Moyal:1949sk}
\begin{equation}
\tilde w(u,v) = \langle e^{-i u \hat x - i v \hat p} \rangle
\end{equation}
whose Taylor series encodes the expectation values \eqref{moments}:
\begin{equation} \label{wtilde}
\tilde w(u,v) = \sum_{m,n \geq 0} \frac{(-i u)^m}{m!} \frac{(-iv)^n}{n!} \langle \sigma( x^m p^n ) \rangle.
\end{equation}
The Wigner function is then defined as the Fourier transform of $\tilde w(u,v)$:
\begin{equation} \label{w}
    w(x,p) = \int \frac{du}{2\pi} \frac{dv}{2 \pi} w(u,v) \; e^{i u x + i v p}.
\end{equation}
The property \eqref{moments} follows directly from \eqref{wtilde} and \eqref{w}.

To find explicit formulae for the characteristic function, we can use the identity
\begin{equation} \label{displacement-nonrelativistic}
    e^{-i u \hat x - i v \hat p} = e^{-i \frac{u}{2} \hat x} e^{- i v \hat p} e^{-i \frac{u}{2} \hat x}.
\end{equation}
to obtain the well-known expressions
\begin{align} \label{wtildefromrho}
    \tilde w (u,v) &= \int dk \, e^{- i v k} \bra{k + \tfrac{u}{2}} \hat \rho \ket{k - \tfrac{u}{2}}, \\
    w(x,p) &= \int \frac{du}{2\pi} e^{i u x} \bra{p + \tfrac{u}{2}} \hat \rho \ket{p - \tfrac{u}{2}}. \label{wfromrho}
\end{align}

To reconstruct $w$ experimentally, one possible method is to measure operators of the form $u \hat x + v \hat p$.
The probability density associated with each such observable gives the integral of the Wigner function over a line in phase space.
The collection of all such line integrals is the Radon transform of $w$, from which $w$ can be recovered by the inverse Radon transform.

Having obtained $w$, or equivalently $\tilde w$, the formula \eqref{wfromrho} can be directly inverted to find the density matrix. 
However it will be instructive to proceed slightly more abstractly.
The formula \eqref{wtilde} can be viewed as a Hilbert-Schmidt inner product between the density operator and the displacement operator:
\begin{equation}
\tilde w(u,v) = \tr ( \hat \rho \, e^{- i u \hat x - i v \hat p}).
\end{equation}
To invert this relation, we can use the fact that the displacement operators form a continuous orthogonal basis for the space of operators:
\begin{equation}
\tr(e^{-i u \hat x - i v \hat p} e^{i u' \hat x + i v' \hat p} ) 
= 2\pi \delta(u - u') \delta (v - v').
\end{equation}
It then follows that $\hat \rho$ can be obtained from $\tilde w$ as
\begin{equation}
\hat \rho = \frac{1}{2 \pi} \int \, du \, dv \, \tilde w(u,v) \, e^{i u \hat x + i v \hat p}.
\end{equation}
Thus to recover the density matrix, it is sufficient to recover the Wigner function

\section{Relativistic Wigner function}

We now turn to the question of whether joint measurements of $\hat K$ and $\hat P$ can be used to recover the density operator $\hat \rho$.
Our procedure follows almost exactly the steps of the previous section: we construct a \emph{relativistic Wigner function} $W(K,P)$ whose Radon transform is directly related to expectation values of linear combinations of $\hat K$ and $\hat P$, and show that $\hat \rho$ can be recovered from it.
The essential difference is that the commutator between $\hat K$ and $\hat P$ \eqref{KPcommutator} differs from that between $\hat x$ and $\hat p$, which reflects the difference between the Lorentz and Galilean groups.

We first define the characteristic function, which encodes the joint moments of $\hat K$ and $\hat P$:
\begin{equation} \label{Wtildefromrho}
    \tilde W(u,v) = \langle e^{-i u \cdot \hat K - i v \cdot \hat P} \rangle.
\end{equation}
Here $u$ and $v$ are $d$-dimensional vectors, with the interpretation of a rapidity and displacement respectively.
The Wigner function is then defined as the Fourier transform,
\begin{equation}
    W(K,P) = \int \frac{d^du}{(2 \pi)^d} \frac{d^d v}{(2 \pi)^d} \, \tilde W(u,v) \, e^{i u K + i v P}.
\end{equation}
Either $W$ or $\tilde W$ can be recovered from measurements of $u \cdot \hat K + v \cdot \hat P$, just as in the nonrelativistic case.

An explicit expression for the characteristic function in momentum space follows from the generalization of \eqref{displacement-nonrelativistic}:
\begin{equation} \label{displacement-relativistic}
e^{-i u \cdot \hat K - i v \cdot \hat P} = e^{- i u \cdot \hat K / 2} e^{-i v \cdot A(u) \cdot \hat P} e^{- i u \cdot \hat K/2}.
\end{equation}
where $A(u)$ is a $d \times d$ matrix with elements
\begin{equation}
A(u)_{ij} = \delta_{ij} + \left(\frac{\sinh(u/2)}{u/2} - 1 \right) \frac{u_i u_j}{u^2}.
\end{equation}
The matrix $A$ reflects the nontrivial transformation of $\hat P$ under boosts, an effect that disappears in the nonrelativistic limit $u \to 0$.

The identity \eqref{displacement-relativistic} leads to the explicit expression
\begin{equation}
    \tilde W(u, v) = \int \widetilde{dk} \; e^{-i v \cdot A(u) \cdot k} \; \rho(\Lambda_{-u/2} k, \Lambda_{u/2} k),
\end{equation}
which is the relativistic generalization of \eqref{wtildefromrho}: the measure is replaced with the relativistically-invariant measure $\widetilde dk$, the Galilean boost $k \to k - u/2$ is replaced by a Lorentz boost $k \to \Lambda_{u/2} k$, and the matrix $A(u)$ appears in the exponent.
Taking the Fourier transform gives the corresponding generalization of \eqref{wfromrho} for $W$
\begin{equation}
W(K,P) = \int \frac{d^d u}{(2 \pi)^{2d}} \frac{e^{i u \cdot K}}{2 \omega_{k^*}} \frac{u/2}{\sinh(u/2)} \rho\left( \Lambda_{-u/2} k^*, \Lambda_{u/2} k^* \right).
\end{equation}
where $k^* := A(u)^{-1} P$.

The essential property we require is that the Wigner function faithfully encodes the state.
Unlike in the nonrelativistic case, the relativistic displacement operators \eqref{displacement-relativistic} are not orthogonal in the Hilbert-Schmidt inner product.
To invert \eqref{Wtildefromrho} to obtain $\hat \rho$ we define the dual operators
\begin{equation}
    \hat{\mathcal{O}}(u,v) = e^{i \, u \cdot \hat K/2} \left( \frac{\sinh(u/2) \hat P^0}{\pi u} \right)^d e^{i \, v \cdot A(u) \cdot \hat P} e^{i \,u \cdot \hat K/2}.
\end{equation}
which are dual to the displacement operators in the sense that
\begin{equation}
    \tr \left( \hat{\mathcal{O}}(u',v') e^{- i u \cdot \hat K - i v \cdot \hat P} \right) = \delta^d(u-u') \delta^d(v-v').
\end{equation}
Using this identity, the density matrix is given by:
\begin{equation} \label{rhofromWtilde}
    \hat \rho = \int d^d u \; d^d v \; \tilde W(u,v) \; \hat{\mathcal{O}}(u,v).
\end{equation}
Thus we see that the characteristic function (and therefore also the Wigner function) gives a faithful encoding of the density operator.

\section{Gravitational tomography}

In summary, we give the following method to determine a density matrix from its asymptotic gravitational field:
\begin{itemize}
\item Collect statistics from measurement of operators of the form $u \cdot \hat K + v \cdot \hat P$.
\item Find the characteristic function $\tilde W(u,v)$ using the inverse Radon transform.
\item Determine $\hat \rho$ from $\tilde W(u,v)$ using \eqref{rhofromWtilde}.
\end{itemize}
This gives an explicit method by which complete quantum information about a single particle can be decoded from the asymptotic metric.

\section{Discussion}

We have shown that the density operator of a single particle state can be recovered from measurement of the gravitational field at spatial infinity.
This provides a proof of principle for a method to extract an infinite amount of quantum information from the asymptotic gravitational field.
Moreover, the reconstruction procedure is explicit and makes no use of quantum gravity beyond the well-understood perturbative regime.
The essential property is the gravitational constraints, which are a reflection of diffeomorphism invariance.

The relativistic Wigner function we have defined may find applications beyond the one considered here.
It would be natural to consider generalizations to particles with spin; in particular a generalization to spin-1 may be useful in characterizing single-photon states.
It may also be possible to generalize to asymptotically de Sitter or anti de Sitter spacetimes, which have a different group of global symmetries.

It would also be interesting to consider more direct methods to measure the Wigner function.
The value of the Wigner function at a point is itself an observable, which in the nonrelativistic case can be expressed as the expectation value of the displaced parity operator \cite{Royer:1977zz}.
This insight has led to efficient methods to directly measure the Wigner function of an electromagnetic mode without resorting to the inverse Radon transform.
We do not know whether a similar expression obtains in the relativistic case, and whether this can lead to a more direct way of accessing the Wigner function by coupling detectors to the gravitational field.

The most significant limitation of this construction is the restriction to single particle states.
This is related to the fact that such states form irreducible representations of the Poincar\'e group. 
A superposition of states with different Casimirs (mass and spin) can therefore not be distinguished by the global charges, since these commute with the Casimirs.

However, it has been known for some time that the Poincar\'e charges are not the only global charges in asymptotically flat spacetime.
The Bondi-van der Burg-Metzner-Sachs group \cite{Bondi:1962px,Sachs:1962wk} extends Poincar\'e by an infinite set of supertranslation charges which describe the momentum density at each point of an asymptotic sphere.
These charges are sufficient to recover the momenta of individual particles in a multiparticle state \cite{Carney:2017jut,Strominger:2017aeh,Carney:2017oxp}.
However the supertranslation charges are insensitive to phase information in the momentum basis, which requires measurement of a complementary observable.
Our work here suggests that a natural complementary observable would be an angle-dependent generalization of the boost charge $\hat K$: in fact such a generalization has been proposed and is known as the superrotation charge \cite{Barnich:2009se,Barnich:2011ct}.
It is an open question whether the supertranslation and superrotation charges together encode complete information about a multiparticle state.
This would provide a natural mechanism by which information is holographically encoded on the boundary of an asymptotically flat spacetime, and a possible holographic resolution of the black hole information loss paradox.

\section*{Acknowledgments}
I am grateful to Steve Giddings, Achim Kempf and Maria Papageorgiou for helpful discussions and comments.
This research was supported in part by Perimeter Institute for Theoretical Physics. 
Research at Perimeter Institute is supported by the Government of Canada through Industry Canada and by the Province of Ontario through the Ministry of Research and Innovation.

\bibliographystyle{utphys}
\bibliography{wigner}

\providecommand{\href}[2]{#2}\begingroup\raggedright\begin{thebibliography}{10}

\bibitem{tHooft:1993dmi}
G.~'t~Hooft, ``{Dimensional reduction in quantum gravity},'' {\em Conf. Proc.}
  {\bfseries C930308} (1993) 284--296,
\href{http://arxiv.org/abs/gr-qc/9310026}{{\ttfamily arXiv:gr-qc/9310026
  [gr-qc]}}.

\bibitem{Susskind:1994vu}
L.~Susskind, ``{The World as a hologram},''
  \href{http://dx.doi.org/10.1063/1.531249}{{\em J. Math. Phys.} {\bfseries 36}
  (1995) 6377--6396},
\href{http://arxiv.org/abs/hep-th/9409089}{{\ttfamily arXiv:hep-th/9409089
  [hep-th]}}.

\bibitem{Arnowitt:1962hi}
R.~L. Arnowitt, S.~Deser, and C.~W. Misner, ``{The Dynamics of general
  relativity},'' \href{http://dx.doi.org/10.1007/s10714-008-0661-1}{{\em Gen.
  Rel. Grav.} {\bfseries 40} (2008) 1997--2027},
\href{http://arxiv.org/abs/gr-qc/0405109}{{\ttfamily arXiv:gr-qc/0405109
  [gr-qc]}}.

\bibitem{Marolf:2008mf}
D.~Marolf, ``{Unitarity and Holography in Gravitational Physics},''
  \href{http://dx.doi.org/10.1103/PhysRevD.79.044010}{{\em Phys. Rev.}
  {\bfseries D79} (2009) 044010},
\href{http://arxiv.org/abs/0808.2842}{{\ttfamily arXiv:0808.2842 [gr-qc]}}.

\bibitem{JACOBSON:2013ewa}
T.~Jacobson, ``{Boundary unitarity and the black hole information paradox},''
  \href{http://dx.doi.org/10.1142/S0218271813420029}{{\em Int. J. Mod. Phys.}
  {\bfseries D22} (2013) 1342002},
\href{http://arxiv.org/abs/1212.6944}{{\ttfamily arXiv:1212.6944 [hep-th]}}.

\bibitem{DeJonckheere:2017qkk}
T.~De~Jonckheere, ``{Modave lectures on bulk reconstruction in AdS/CFT},''
  \href{http://dx.doi.org/10.22323/1.323.0005}{{\em PoS} {\bfseries Modave2017}
  (2018) 005},
\href{http://arxiv.org/abs/1711.07787}{{\ttfamily arXiv:1711.07787 [hep-th]}}.

\bibitem{Harlow:2018fse}
D.~Harlow, ``{TASI Lectures on the Emergence of the Bulk in AdS/CFT},''
\href{http://arxiv.org/abs/1802.01040}{{\ttfamily arXiv:1802.01040 [hep-th]}}.

\bibitem{Hamilton:2006az}
A.~Hamilton, D.~N. Kabat, G.~Lifschytz, and D.~A. Lowe, ``{Holographic
  representation of local bulk operators},''
  \href{http://dx.doi.org/10.1103/PhysRevD.74.066009}{{\em Phys. Rev.}
  {\bfseries D74} (2006) 066009},
\href{http://arxiv.org/abs/hep-th/0606141}{{\ttfamily arXiv:hep-th/0606141
  [hep-th]}}.

\bibitem{Balasubramanian:2006iw}
V.~Balasubramanian, D.~Marolf, and M.~Rozali, ``{Information Recovery From
  Black Holes},'' \href{http://dx.doi.org/10.1007/s10714-006-0344-8,
  10.1142/S0218271806009765}{{\em Gen. Rel. Grav.} {\bfseries 38} (2006)
  1529--1536}, \href{http://arxiv.org/abs/hep-th/0604045}{{\ttfamily
  arXiv:hep-th/0604045 [hep-th]}}.
[Int. J. Mod. Phys.D15,2285(2006)].

\bibitem{Regge:1974zd}
T.~Regge and C.~Teitelboim, ``{Role of Surface Integrals in the Hamiltonian
  Formulation of General Relativity},''
\href{http://dx.doi.org/10.1016/0003-4916(74)90404-7}{{\em Annals Phys.}
  {\bfseries 88} (1974) 286}.

\bibitem{Chatwin-Davies:2015hna}
A.~Chatwin-Davies, A.~S. Jermyn, and S.~M. Carroll, ``{How to Recover a Qubit
  That Has Fallen Into a Black Hole},''
  \href{http://dx.doi.org/10.1103/PhysRevLett.115.261302}{{\em Phys. Rev.
  Lett.} {\bfseries 115} no.~26, (2015) 261302},
\href{http://arxiv.org/abs/1507.03592}{{\ttfamily arXiv:1507.03592 [hep-th]}}.

\bibitem{Srednicki:2007qs}
M.~Srednicki, {\em {Quantum field theory}}.
\newblock Cambridge University Press,
2007.
\newblock

\bibitem{Feynman:1963ax}
R.~P. Feynman, ``{Quantum theory of gravitation},'' {\em Acta Phys. Polon.}
  {\bfseries 24} (1963) 697--722.
[,272(1963)].

\bibitem{Donnelly:2015hta}
W.~Donnelly and S.~B. Giddings, ``{Diffeomorphism-invariant observables and
  their nonlocal algebra},''
  \href{http://dx.doi.org/10.1103/PhysRevD.94.029903,
  10.1103/PhysRevD.93.024030}{{\em Phys. Rev.} {\bfseries D93} no.~2, (2016)
  024030}, \href{http://arxiv.org/abs/1507.07921}{{\ttfamily arXiv:1507.07921
  [hep-th]}}.
[Erratum: Phys. Rev.D94,no.2,029903(2016)].

\bibitem{Donnelly:2016rvo}
W.~Donnelly and S.~B. Giddings, ``{Observables, gravitational dressing, and
  obstructions to locality and subsystems},''
  \href{http://dx.doi.org/10.1103/PhysRevD.94.104038}{{\em Phys. Rev.}
  {\bfseries D94} no.~10, (2016) 104038},
\href{http://arxiv.org/abs/1607.01025}{{\ttfamily arXiv:1607.01025 [hep-th]}}.

\bibitem{Donnelly:2017jcd}
W.~Donnelly and S.~B. Giddings, ``{How is quantum information localized in
  gravity?},'' \href{http://dx.doi.org/10.1103/PhysRevD.96.086013}{{\em Phys.
  Rev.} {\bfseries D96} no.~8, (2017) 086013},
\href{http://arxiv.org/abs/1706.03104}{{\ttfamily arXiv:1706.03104 [hep-th]}}.

\bibitem{Donnelly:2018nbv}
W.~Donnelly and S.~B. Giddings, ``{Gravitational splitting at first-order:
  quantum information localization in gravity},''
\href{http://arxiv.org/abs/1805.11095}{{\ttfamily arXiv:1805.11095 [hep-th]}}.

\bibitem{Wigner:1932eb}
E.~P. Wigner, ``{On the quantum correction for thermodynamic equilibrium},''
\href{http://dx.doi.org/10.1103/PhysRev.40.749}{{\em Phys. Rev.} {\bfseries 40}
  (1932) 749--760}.

\bibitem{Moyal:1949sk}
J.~E. Moyal, ``{Quantum mechanics as a statistical theory},''
\href{http://dx.doi.org/10.1017/S0305004100000487}{{\em Proc. Cambridge Phil.
  Soc.} {\bfseries 45} (1949) 99--124}.

\bibitem{Royer:1977zz}
A.~Royer, ``{Wigner function as the expectation value of a parity operator},''
\href{http://dx.doi.org/10.1103/PhysRevA.15.449}{{\em Phys. Rev.} {\bfseries
  A15} (1977) 449--450}.

\bibitem{Bondi:1962px}
H.~Bondi, M.~G.~J. van~der Burg, and A.~W.~K. Metzner, ``{Gravitational waves
  in general relativity. 7. Waves from axisymmetric isolated systems},''
\href{http://dx.doi.org/10.1098/rspa.1962.0161}{{\em Proc. Roy. Soc. Lond.}
  {\bfseries A269} (1962) 21--52}.

\bibitem{Sachs:1962wk}
R.~K. Sachs, ``{Gravitational waves in general relativity. 8. Waves in
  asymptotically flat space-times},''
\href{http://dx.doi.org/10.1098/rspa.1962.0206}{{\em Proc. Roy. Soc. Lond.}
  {\bfseries A270} (1962) 103--126}.

\bibitem{Carney:2017jut}
D.~Carney, L.~Chaurette, D.~Neuenfeld, and G.~W. Semenoff, ``{Infrared quantum
  information},'' \href{http://dx.doi.org/10.1103/PhysRevLett.119.180502}{{\em
  Phys. Rev. Lett.} {\bfseries 119} no.~18, (2017) 180502},
\href{http://arxiv.org/abs/1706.03782}{{\ttfamily arXiv:1706.03782 [hep-th]}}.

\bibitem{Strominger:2017aeh}
A.~Strominger, ``{Black Hole Information Revisited},''
\href{http://arxiv.org/abs/1706.07143}{{\ttfamily arXiv:1706.07143 [hep-th]}}.

\bibitem{Carney:2017oxp}
D.~Carney, L.~Chaurette, D.~Neuenfeld, and G.~W. Semenoff, ``{Dressed infrared
  quantum information},''
  \href{http://dx.doi.org/10.1103/PhysRevD.97.025007}{{\em Phys. Rev.}
  {\bfseries D97} no.~2, (2018) 025007},
\href{http://arxiv.org/abs/1710.02531}{{\ttfamily arXiv:1710.02531 [hep-th]}}.

\bibitem{Barnich:2009se}
G.~Barnich and C.~Troessaert, ``{Symmetries of asymptotically flat 4
  dimensional spacetimes at null infinity revisited},''
  \href{http://dx.doi.org/10.1103/PhysRevLett.105.111103}{{\em Phys. Rev.
  Lett.} {\bfseries 105} (2010) 111103},
\href{http://arxiv.org/abs/0909.2617}{{\ttfamily arXiv:0909.2617 [gr-qc]}}.

\bibitem{Barnich:2011ct}
G.~Barnich and C.~Troessaert, ``{Supertranslations call for superrotations},''
  {\em PoS} {\bfseries CNCFG2010} (2010) 010,
  \href{http://arxiv.org/abs/1102.4632}{{\ttfamily arXiv:1102.4632 [gr-qc]}}.
[Ann. U. Craiova Phys.21,S11(2011)].

\end{thebibliography}\endgroup

\end{document}